\documentclass[aps,preprint,preprintnumbers,amsmath,amssymb]{revtex4}
\usepackage{amssymb}
\usepackage{graphicx}
\usepackage{dcolumn}
\usepackage{bm}

\usepackage{color}
\usepackage[colorlinks=true,linkcolor=blue,urlcolor=blue,citecolor=blue]{hyperref}

\makeatletter

\newcommand{\Rmnum}[1]{\expandafter\@slowromancap\romannumeral #1@}

\begin{document}

\title{Pressure induced superconductivity bordering a charge-density-wave state in NbTe$_4$ with strong spin-orbit coupling }

\author{Xiaojun Yang$^{1,2}$, Yonghui Zhou$^3$, Mengmeng Wang$^1$, Hua Bai$^{1}$, Xuliang Chen$^3$, Chao An$^3$, Ying Zhou$^3$, Qian Chen$^{1}$, Yupeng Li$^{1}$, Zhen Wang$^{1}$, Jian Chen$^{1}$, Chao Cao$^{4}$, Yuke Li$^{4}$, Yi Zhou$^{1,6}$, Zhaorong Yang$^3$\footnote[1]{Electronic address: zryang@issp.ac.cn} and Zhu-An Xu$^{1,5,6}$\footnote[2]{Electronic address: zhuan@zju.edu.cn}}
\affiliation{$^{1}$ State Key Laboratory of Silicon Materials and Department of Physics, Zhejiang University, Hangzhou 310027, China.\\
$^2$ School of Physics and Optoelectronics, Xiangtan University, Xiangtan 411105, China\\
$^3$ Anhui Province Key Laboratory of Condensed Matter Physics at Extreme Conditions, High Magnetic Field Laboratory, Chinese Academy of Science, Hefei 230031, China\\
$^4$ Department of Physics, Hangzhou Normal University, Hangzhou 310036, China\\
$^5$ Zhejiang California International NanoSystems Institute,
Zhejiang University, Hangzhou 310058, China\\
$^6$ Collaborative Innovation Centre of Advanced Microstructures,
Nanjing 210093, P. R. China}

\date{\today}
\maketitle

\textbf{ Transition-metal chalcogenides host various phases of
matter, such as charge-density wave (CDW), superconductors, and
topological insulators or semimetals. Superconductivity and its
competition with CDW in low-dimensional compounds have attracted
much interest and stimulated considerable research. Here we report
pressure induced superconductivity in a strong spin-orbit (SO)
coupled quasi-one-dimensional (1D) transition-metal chalcogenide
NbTe$_4$, which is a CDW material under ambient pressure. With
increasing pressure, the CDW transition temperature is gradually
suppressed, and superconducting transition, which is fingerprinted
by a steep resistivity drop, emerges at pressures above 12.4 GPa.
Under pressure $p$ = 69 GPa, zero resistance is detected with a
transition temperature $T_c$ = 2.2 K and an upper critical field
$\mu_0H_{c2}$ = 2 T. We also find large magnetoresistance (MR) up to
102\% at low temperatures, which is a distinct feature
differentiating NbTe$_4$ from other conventional CDW materials.
}\\

\section*{Introduction}

Transition-metal chalcogenides possess 
rich structural chemistry and a wide variety of unusual physical
properties\cite{NbIrTe4_JACS,review_AP,SDW_NbSe2}. 
The latter includes, for instance, charge density
wave (CDW)\cite{SDW_NbSe2}, superconductivity\cite{SC_NbSe2,
SC_TMD_science,WTe2_SC_arXiv1,WTe2_SC_arXiv2} and recently reported
extremely large magnetoresistance in
WTe$_2$\cite{WTe2_nature,WTe2_PRL_Zhu,WTe2_arXiv_SOC,WTe2_arXiv_pressure,WTe2_arXiv_LMR,WTe2_PRL_ARPES}.
Among chalcogenides, tellurides are usually different from sulfides
and selenides in crystal structures, electronic structures and
physical properties, due to the diffusive nature of the tellurium
valence orbitals\cite{Ta4Pd3Te16_JACS} and thus more covalent character
of tellurium\cite{NbIrTe4_JACS}. While sulfides and
selenides, such as NbS$_2$, NbSe$_2$ and NbSe$_3$, were intensively
studied in the context of CDW and/or
superconductivity\cite{SDW_NbSe2,SC_NbSe2,NbSe3_PRB}, tellurides
have not received much attention until
recently\cite{WTe2_nature,WTe2_PRL_Zhu,WTe2_arXiv_SOC,WTe2_arXiv_pressure,WTe2_arXiv_LMR,WTe2_PRL_ARPES}.
One import feature of tellurides is that the atomic number of Te
is very large, resulting in a strong spin-orbital (SO) coupling.
Nowadays, topological materials with strong SO coupling have been drawing plenty of attention in
condensed matter physics\cite{IrTe2_PRL}. It is highly desirable
to discover noval superconductors with strong SO coupling for
understanding the nature of topological superconductivity.
On the other hand, low dimensionality 
is accompanied by strong lattice instability. 
Additional interest for pursuing a quasi-1D material
with itinerant electrons comes from the possible realization of
Luttinger liquid, in which an exotic spin-charge separation is
expected\cite{1D FL_RPP}. 
Thus, it should be of great interest to study superconductivity
in quasi-1D tellurides with large atomic number\cite{Ta4Pd3Te16_JACS,KCr2As2_PRX},
where competing interactions might give rise to interacting ground states.

The magnetoresistance (MR) in ordinary non-magnetic metals is a
relatively weak effect and usually at the level of a few
percent\cite{weak_effect_book,week_effect_SrFBiS2}. Materials
exhibiting large MR are not only utilized to enlarge the
sensitivity of read/write heads of magnetic storage devices, e.g.,
magnetic memory\cite{LMR_memory} and hard
drives\cite{LMR_hard_driver}, but also stimulating many fundamental
researches\cite{LMR_funderment,WTe2_nature}. Typically, large negative MR
occurs in thin-film metals \cite{GMR_JAP}, manganese based
perovskites\cite{CMR_JAP,CMR_Nature} and some disordered
systems\cite{LMR_EPL_Me,LMR_JAP_Me}, while large positive MR has been
observed in semiconductors\cite{AgSe_Nature} and
semimetals\cite{Bi_PR, WTe2_nature}. In general, there exists only a few of CDW materials,
which show large positive MR\cite{NbSe3_LMR_FC,
NbSe3_jpcm_SJQ, AMoO_LMR_PRB}. The origin of the huge positive MR
effect in the CDW state is still under debate. Exploring more CDW
materials with large MR may help to clarify the controversial explanations.

NbTe$_4$ and TaTe$_4$ belong to the same group of quasi-1D CDW
materials (space group P4/mcc). The structure of NbTe$_4$ was first determined by Selte
and Kjekshus in 1964\cite{NbTe4_structure}. The metal atoms Nb
form linear chains along the tetragonal $c$-axis and the Te atoms
form square antiprismatic formulae in which the metal atoms
confined (Fig. 1(a,b)). Superlattice reflections indicate that the
$a$-axis is doubled and the $c$-axis is tripled, leading to an enlarged unit cell 2a* $\times$ 2a*
$\times$ 3c*\cite{NbTe4_structure,NbTe4_modu_1}. NbTe$_4$ undergoes a
strong lattice distortion around room temperature to form an incommensurate
charge density wave (CDW)  phase\cite{NbTe4_modu_1,NbTe4_modu_2,NbTe4_modu_PRB}.
 The resulting CDW superstructure in NbTe$_4$ was visualized by
 scanning tunneling microscopy (STM)\cite{NbTe4_modu_PRB}.
 Intriguingly, the CDW ordering in NbTe$_4$ is highly anisotropic
 due to the quasi-1D chain structure. Strikingly, there exist three CDW orders in NbTe$_4$,
two are incommensurate with wave vectors $\overrightarrow{q}_1$ =
(0, 0, 0.311c*) and $\overrightarrow{q}_2$ = (0.5a*, 0.5b*,
0.344c*), respectively, and the third is commensurate with
$\overrightarrow{q}_3$ = (0.5a*, 0, $\frac{1}{3}$c*)
\cite{NbTe4_modu_1}. The crystal structure of NbTe$_4$ is depicted
in Fig. 1(a, b). The displacements of Nb ion in a single column
for the commensurate phase are shown in Fig. 1c, which can also be
found in previous literature\cite{NbTe4_Lat_distortion}. Indeed,
NbTe$_4$ and TaTe$_4$ are the only two reported crystals in which
three CDWs coexist \cite{NbTe4_modu_1}. In this paper, we report
large magneto-resistance and high pressure induced
superconductivity in NbTe$_4$. The CDW transition temperature is
strongly suppressed by applied pressure, and superconductivity
fingerprint of steep resistivity drop emerges when pressure
exceeds 12.4 GPa. Under pressure $p$ = 69 GPa, zero resistance is
reached with a transition temperature $T_c$ = 2.2 K and an upper
critical field of 2 T. We also observed large magnetoresistance
(MR) up to 102\% at low temperatures in NbTe$_4$, which is rarely
observed in conventional CDW systems.

\section*{Results and Discussions}


Figure 1d displays the X-ray diffraction pattern of NbTe$_4$
single crystal. Only multiple reflections of the (0 $l$ 0) planes can
be detected, consistent with the quasi-one-dimensional crystal
structure depicted in Fig. 1(a, b). The interplane spacing is
determined to be 6.499 $\text{{\AA}}$, agreeing with the
previous reported value of the NbTe$_4$
phase \cite{NbTe4_structure}. For the (030) peak, the full width at half-maximum is only 0.03$^{\circ}$, indicating high quality of the crystals.

The temperature dependence of resistivity under various applied
magnetic fields ($\mu_0H$  up to 15 T) is summarized in Fig. 2a. In
zero field, the room temperature resistivity is 59.2 $\mu\Omega\cdot$cm
and decreases to 9.4 $\mu\Omega\cdot$cm at $T$ = 2 K, yielding a
residual resistivity ratio (RRR) of 6.3. Low RRR in NbTe$_4$ single crystals has been commonly
reported, which is irrespective of growth conditions \cite{NbTe4_R1,NbTe4_R2}.  To understand the origin
of the poor RRR, we performed the measurements of energy
dispersive X-ray spectroscopy (EDXS). The chemical composition
determined by EDXS gives the atomic ratio Nb:Te = 0.8:4,
with a measurement error of $\pm$3.5\% depending on the
elements measured. This result indicates that there exists significant
deficiency for Nb, which may be responsible for the poor RRR. To
get further insight into the resistivity data, we take the partial
derivative of resistivity with respect to temperature, as shown in
the inset of Fig. 2a. At $T^*$ = 200 K a cusp is observed,
consistent with the previous reports\cite{NbTe4_R1,NbTe4_R2}. It is known that the two
incommensurate superlattices with $\overrightarrow{q}_1$ = (0,
0,0.311c*) and $\overrightarrow{q}_2$ = (0.5a*, 0.5b*, 0.344c*)
are stable at room temperature in NbTe$_4$. Below room
temperature, additional superlattice ordering with $\overrightarrow{q}_3$ =
(0.5a*, 0, $\frac{1}{3}$c*) ermerges at about 200 K. The change of the slope in the
temperature dependence of resistivity at 200 K should be
due to the appearance of the $\overrightarrow{q}_3$
superlattice\cite{NbTe4_R1,NbTe4_R2}. Usually, more drastic
anomalies in resistivity should be observed at the CDW transition
temperatures. For example, two sharp increases in resistivity were
observed in NbSe$_3$ at 145 and 59 K \cite{NbSe3_PRB},
corresponding to about 20\% and 48\% of conduction electrons
condensating into the CDW states,
respectively\cite{NbSe3_PRB,NbSe3_jpcm_SJQ}. The observed small
change of resistivity around 200 K in NbTe$_4$ is indicative of a
finite but slight decrease of free carrier density associated with
the reconstruction of the Fermi surface by Brillouin zone
refolding.
The anomaly at around $T_L$ = 50 K may be due to the
lock-in transition into the 2a* $\times$ 2a* $\times$ 3c*
superstructure, which is worthy of further clarification\cite{NbTe4_R2}.
The anomies at $T^*$ = 200 K
and $T_L$ = 50 K, which were also observed in previous
reports\cite{NbTe4_R1,NbTe4_R2}, further confirm the CDW
features observed in NbTe$_4$.

The field ($H$//$b$) dependence of resistance ($I$//$c$) at various temperatures is
shown in Fig. 2b. The magneto-resistance (MR), which is defined
as MR = $[\rho(15 T)-\rho(0)]/\rho(0)$, rises up to 102\% under
a magnetic field of 15 T at $T$ = 2 K. Although the MR value of NbTe$_4$
is orders of magnitude smaller than that of WTe$_2$\cite{WTe2_nature} and
Bi\cite{Bi_PR}, it is much larger than that of a usual single-band
weakly interacting electron system, in which the MR is usually at the level of a
few percent\cite{weak_effect_book,WTe2_nature}. For a single-band  noninteracting electron
system, the Hall field exactly balances the Lorentz force, and the
electron moves as if in zero field without being deflected; thus,
there is no remarkable magnetoresistance\cite{weak_effect_book,MgB2_PRL}.

Figure 3 displays the field ($H$//$b$) dependence of Hall resistivity ($I$//$c$)
$\rho_{yx}$ at various temperatures. At 2 K, $\rho_{yx}$ is
positive under low fields but switches to negative sign in
higher fields. With increasing temperature, the required field where
$\rho_{yx}$ changes its sign decreases. The curvature and sign
reversal in the Hall resistivity indicate the coexistence of
hole-type minority carriers with high mobility and electron-type
majority carriers with low mobility\cite{EuFBiS_PRB,NbSb2_SR}.
The multiband nature of NbTe$_4$ is also manifested in the breakdown of the
Kohler's rule, as plotted in the inset of Fig. 3. According to the
Kohler's rule, if only one relaxation time $\tau$ exists in
metals, then MR can be characterized by a function of
$\mu_0H/\rho(0)$, and the results for different temperatures
should collapse into a single curve\cite{MgB2_PRL, Oxford_book}.
As shown clearly in the inset of Fig. 3, the Kohler's rule is
violated in NbTe$_4$, as evident by the non-overlapping of the
MR curves at different temperatures. The observeation indicates more than
one relaxation time $\tau$ exist, supporting the multiband result
obtained by the Hall measurements.

Materials exhibiting a large magneto-resistance have potential
device applications and thus have been attracting researchers' interest
\cite{PdCoO2_PRL}. The most well-known examples are the giant
magnetoresistance in magnetic multilayers\cite{GMR_JAP} and
colossal magnetoresistance in manganites\cite{CMR_JAP}, both of
them rely on the coupling between spin configuration and charge
transport. However, even among nonmagnetic materials, extremely large magnetoresistance
(MR) may arise in semimetals with electron-hole Fermi surface
(FS) compensation\cite{FC_science_Bi,APL_PtBi2,FC_PRL_Bi_and_C,FC_PRL_URu2Si2}.

As a discussion of the multiband effects in NbTe$_4$,
we define the direction of the current (Hall
voltage) as the $x$-axis ($y$-axis). Even though no net current
should exist in the $y$-direction, the currents in the
$y$-direction by one particular type of carriers may be non-zero in a
multiband system. When we applied magnetic field, the
$y$-direction currents should be affected by the Lorentz force which
is antiparallel to the $x$-direction\cite{FC_book}. The back flow
of carriers provides a substantial source of the large
magnetoresistance in metals with multiple bands like MgB$_2$\cite{MgB2_PRL} and semimetals like Bi\cite{FC_PRL_Bi_and_C} and WTe$_2$\cite{WTe2_nature}.
Owing to the coexistence of both
electron- and hole-type carriers, the large MR in NbTe$_4$ may be
attributed to multiband effects. 
We employed a two band mode to fit the $\rho_{xx}$ and $\rho_{yx}$
data simultaneously in the low field region\cite{SI}. The values
of $n_e$ and $n_h$ are close to each other at all the measured
temperatures, which suggests that the large MR should result from
the electron-hole compensation effect. The mobilities $\mu_e$ and
$\mu_h$ increase with decreasing temperatures as the usual metals.
Meanwhile, the carrier concentrations decrease significantly at
low temperatures, which occurs commonly in a CDW system. At $T$ =
2 K, $\mu_e$ is 0.22 m$^2$V$^{-1}$s$^{-1}$, while $\mu_h$ is 0.29
m$^2$V$^{-1}$s$^{-1}$. Actually large magnetoresistance was also
observed in the CDW materials NbSe$_3$\cite{NbSe3_LMR_FC,
NbSe3_jpcm_SJQ} and AMo$_6$O$_{17}$ (A = Na, K, and
Tl)\cite{AMoO_LMR_PRB}. It has been proposed that the large MR in
these CDW systems may result from the magnetic-field-induced
enhancement of the CDW gap\cite{NbSe3_LMR_FC}, but a study by
Tritt et al.\cite{NbSe3_CDW_gap_in} presented negative results on
such a claim. Until now, the nature of the huge positive MR effect
in the CDW state is still ambiguous.  Actually, only a few CDW
materials show large positive MR\cite{NbSe3_LMR_FC,
NbSe3_jpcm_SJQ, AMoO_LMR_PRB}. The origin of the large MR in
NbTe$_4$ deserves further investigation.

Superconductivity often occurs in the proximity of other competing ordered states.
Both high-$T_c$ cuprates and Fe-based superconductors  are close
to antiferromagnetic (AFM)
ordered states\cite{LaFeAsO_order_Nature,LaFeAsO_order_NP}. The AFM order
in Fe-based superconductor is of a spin density wave (SDW)-type.
For CDW materials, high pressure or chemical doping can
continuously suppress the CDW order, and then superconducting
transition temperature is enhanced or superconducting state
emerges after the suppression of CDW
state\cite{CuTiSe2_Cava,NbSe2_PRL}. The weakening of competing
orders normally favors superconductivity\cite{NbSe3_EPL_CDW}. The CDW order
of NbTe$_4$ survives up to very high temperatures. Thus it is
interesting to investigate whether superconductivity can be
induced by pressure.

Figure 4a shows the evolution of resistance ($I//c$) as a function
of temperature of the NbTe$_4$ single crystal at various pressures
from 5 GPa to 69 GPa. The samples used in high pressure and
ambient pressure measurements are two different ones, but they are
from the same batch. Under an applied pressure of $p$ = 5 GPa, the
cusp associated with the appearance of the
$\overrightarrow{q}_3$-superlattice CDW transition is suppressed
to $T^*$ = 173 K, as shown in Fig. 4b. The MR decreases to less
than 10\% with an applied pressure of 5 GPa, and becomes smaller
and smaller with an increasing pressure, as shown in Fig. 4c. When
the pressure increases up to 12.4 GPa, the cusp associated with
the CDW transition is suppressed and becomes too weak to be
distinguished in the resistivity curve, instead, a sudden
resistivity decrease presents at $T$ $\sim$ 2.4 K, which could be
a fingerprint of superconductivity. So we keep increasing the
pressure to trace the superconducting transition. Upon further
increasing pressure, at $T$ = 1.7 K, which is the base temperature
of our high pressure measurement system, the resistance drops to
very small. Finally, under $p$ = 69 GPa, which is the highest
pressure we can apply, zero resistance has been detected. The
low-T resistance appears to increase monotonically with pressure.
In contrast, the room-temperature resistance keeps increasing up
to 12.4 GPa, and then steadily decreases, which coincides with the
disappearance of the CDW phase. Whether the variation in
room-temperature resistance is correlated with the CDW order
deserves further clarification. T$_c$ around 2.2 K is stabilized
for the pressure range from 10 to 69 GPa. Such a wide
stabilization pressure range could be a result of heavy doping due
to Nb deficiency up to 0.2. Fig. 4d shows the suppression of
superconductivity under magnetic fields ($H//b$). The inset gives
the temperature dependence of the upper critical field
$\mu_0H_{c2}(T)$, determined by using 90\% normal state
resistivity criterion. The temperature dependence of
$\mu_0H_{c2}(T)$ is nearly linear in the investigated temperature
range. According to the Ginzburg-Landau theory, the upper critical
field $H_{c2}$ evolves with temperature following the formula
$H_{c2}(T) = H_{c2}(0)(1 - t^2)/(1 + t^2)$, where $t$ is the
renormalized temperature $T/T_c$. It is found that the
experimental upper critical field $\mu_0H_{c2}(T)$ can be well
fitted by this model and its zero-temperature limit is extracted
to be 2.0 T. Due to the limited temperature range, the estimated
$\mu_0H_{c2}(0)$ by this way may be of considerable error.


The temperature-pressure phase diagram is summarized in Fig. 5.
The CDW transition temperature in NbTe$_4$ is strongly suppressed
with an applied high pressure. Accompanying the suppression of the CDW order, superconductivity fingerprint appears. Such
superconducting phase diagrams, where superconductivity may
compete with different kind of orders, have been observed in many
systems, including high-$T_c$ cuprates, Fe-based superconductors,
NbSe$_2$ and Cu$_x$TiSe$_2$\cite{LaFeAsO_order_Nature,LaFeAsO_order_NP,NbSe2_PRL,CuTiSe2_Cava}.
Barath $et$ $al.$ proposed that the superconducting paring mechanism
in Cu$_x$TiSe$_2$ could be associated with the quantum criticality,
stemming from the fluctuations of CDW order\cite{CuTiSe2_PRL_fluctuation}.
There are also many theoretical works on
the superconducting dome in transition metal dichalcogenides (TMDCs),
and pursuing the nature of the observed superconducting states is still on-going\cite{TMD_PRB_Das,TMD_PRB_Rosner}.
High temperature high pressure treatment could induce phase
transformation in NbTe$_4$\cite{NbTe4_HighPressure}, which may be
relevant to the occurrence of superconductivity. Further
investigation on this issue is of potential interest. Since the
physical properties of NbTe$_4$ and TaTe$_4$ show numerous similar
features, the further research on pressure effect of TaTe$_4$
should be gainful\cite{TaTe4_APL_MR}. 
As a low dimensional
chalcogenide, it could be fascinating to explore whether 
intercalation or doping could induce superconductivity, or at
least decrease the critical superconducting pressure, as the case
in Cu$_x$TiSe$_2$\cite{CuTiSe2_PRL_fluctuation},
Cu$_x$Bi$_2$Se$_3$\cite{CuBi2Se3_PRL_Cava}, Ir$_{1-x}$
Pd$_x$Te$_2$\cite{IrTe2_PRL},
(NbSe$_4$)$_{3.33}$I\cite{(NbSe4)3.33I_SST_Wang}. Primarily
because of the strong SO coupling, the appealing Majorana surface
state has been suggested to exist in Cu$_x$Bi$_2$Se$_3$
superconductor \cite{CuBi2Se3_PRL_Cava,CuBi2Se3_PRL_Cava_Fu}. Due
to the large atom number in NbTe$_4$, it is expected to own strong
SO coupling. Furthermore, a recent theoretical work has proposed
monolayer hole-doped TMDCs as candidates for topological
superconductors\cite{TMDs_NC_Hsu}, which makes transition metal
chalcogenide systems more fascinating. Thus it becomes quite
interesting to explore the possible presence of nonconventional
quantum states in NbTe$_4$. Our work may provide another promising
system for exploring novel topological superconductors.

\section*{Summary}

In conclusion, we discovered unusually large magnetoresistance(MR)
and pressure induced superconductivity with a maximum $T_c$ of 2.2
K in a quasi-one-dimensional (1D) transition-metal chalcogenide
NbTe$_4$ with strong spin-orbit coupling. Superconductivity
appears when the CDW order is suppressed by high pressure.
Although $T_c$ is relatively low, the large MR and strong SO
coupling make NbTe$_4$ a promising candidate for the exploration
of novel superconductors.

\section*{Methods}

\textbf{Sample synthesis and characterization}

The NbTe$_4$ crystals were grown by a self-flux method.
Powders of the elements Nb (99.97\%) and Te (99.99\%), all from Alfa Aesar, in an atomic
ratio of Nb:Te = 1:8 were thoroughly mixed together,
loaded, and sealed into an evacuated quartz ampule. The
ampule was slowly heated up to 1273 K and held for 25 h. After
that, it was slowly cooled to 873 K at a rate of 3 K/h, followed
by furnace cooling down to room temperature. Shiny, gray-black soft
crystals in flattened needle shapes were harvested with a typical
dimension of 1.2 $\times$ 0.02 $\times$ 0.02 mm$^3$, as shown in Fig. 1(e).
Single crystal X-ray diffraction (XRD) was performed at room temperature
using a PANalytical X-ray diffractometer (Model EMPYREAN) with a monochromatic
CuK$_{\alpha1}$ radiation. Energy-dispersive x-ray
spectroscopy (EDXS) was collected by an Octane Plus Detector
(AMETEX EDAX), equipped in a field-emitting
scanning electron microscope (SEM, Hitachi S-4800).

\textbf{Measurements}

The resistance data was collected using standard four-probe method in a
screw-pressure-type diamond anvil
cell (DAC), which is made of non-magnetic Cu-Be alloy. The diamond culet was about 320 $\mu$m in
diameter. A T301 stainless steel gasket was pre-indented from a thickness of 220 $\mu$m
to 35 $\mu$m, leaving a pit inside the gasket. A hole with diameter of 300 $\mu$m was drilled
in the center of the pit by laser ablation. The pit of the indented gasket was then
covered with a mixture of epoxy and fine cubic boron nitride (cBN) powder and
compressed tightly to insulate the electrode leads from the metallic gasket. Next, the
cBN-covered pit served as the sample chamber, where a NbTe$_4$ single crystal in
dimension of 200 $\mu$m $\times$ 35 $\mu$m $\times$ 5 $\mu$m was inserted without the pressure-transmitting
medium, together with a ruby ball served as a pressure marker at the top of the
sample. The value of the pressure was determined by the ruby fluorescence method. Platinum (Pt)
foil with a thickness of 5 $\mu$m was used as electrodes. The gasket surface outside the
pit was insulated from the electrode leads by a layer of Scotch tape. The DAC was
put inside a home-made multifunctional measurement system (1.8-300 K, JANIS
Research Company Inc.; 0-9 T, Cryomagnetics Inc.) with helium (He) as the medium
for heat conduction to obtain high efficiency of heat transfer and good precision of
temperature control. Two Cernox resistors (CX-1050-CU-HT-1.4 L) located near the
DAC were employed to ensure the accuracy of temperature in the presence of
magnetic field. The ambient-pressure electrical transport measurements were carried out in a
Oxford cryostat system with magnetic field up to 15 T and temperature down to 1.5 K.
Ohmic contacts were made with gold wires and silver paste.

\section*{Acknowledgements}

We thank Yi Zheng and Guanghan Cao for valuable discussions and
helpful suggestions. This work is supported by the Ministry of
Science and Technology of China (Grant No. 2016YFA0300402 and
2016YFA0401804), NSF of China (Contract Nos. U1332209, 11574323,
and U1632275). X. C. thanks for the support from the Natural
Science Foundation of Anhui Province (1708085QA19).


\section*{Author contributions}

Zhu-an Xu and Xiaojun Yang designed the research. Zhu-an Xu
administered the experiment. Xiaojun Yang synthesized the samples.
Xiaojun Yang and Yonghui Zhou performed the measurements. Mengmeng
Wang, Hua Bai, Xuliang Chen, Chao An, Ying Zhou, Qian Chen, Yupeng
Li, Zhen Wang, Jian Chen, Chao Cao, Yuke Li, Yi Zhou and
Zhaorong Yang presented helpful suggestions in the measurements
and in the data analysis. Zhu-an Xu, Xiaojun Yang and Zhaorong
Yang analyzed the data, interpreted the results, and wrote the
paper.

\section*{Additional Information}
\textbf{Competing Interests:} The authors declare no competing
interests.

Correspondence and requests for materials should be addressed to
Zhu-an Xu (zhuan@zju.edu.cn) and Zhaorong Yang
(zryang@issp.ac.cn).

\newpage
\textbf{Figures:}

\begin{figure}[!h]
\includegraphics[width=0.5\columnwidth]{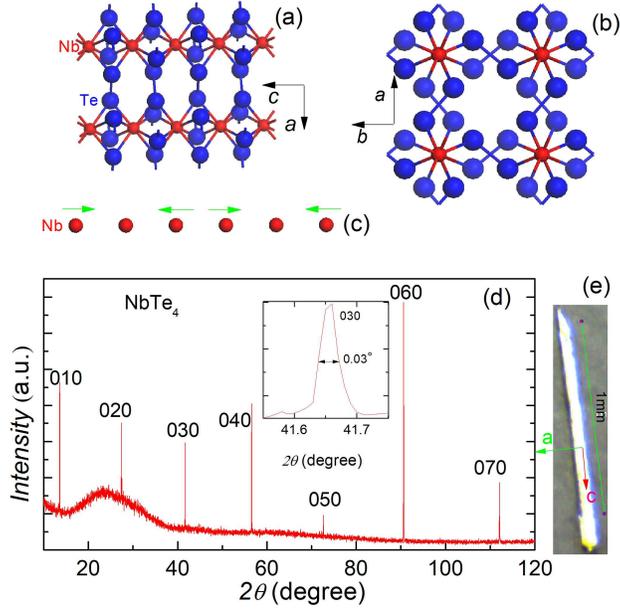}
\caption{\label{Fig. 1} \textbf{Structural characterization by
x-ray diffractions for NbTe$_4$.} \textbf{(a,b)}, The basic
crystal structure of NbTe$_4$ projected along the [010]
\textbf{(a)} and [001] \textbf{(b)} directions. \textbf{(c)}, The
red solid circles represent equidistant Nb ions on the axis of a
single column of the basic structure; the arrows show the
displacements and resultant trimerizations of the Nb ions in the
commensurate phase. \textbf{(d)}, XRD structure characterization
of a NbTe$_4$ single crystal. Only (0 $l$ 0) peaks can be
observed. The inset: The enlarged XRD curve of (030) peak.
\textbf{(e)}, A typical crystal of NbTe$_4$, with crystallographic
directions marked.}
\end{figure}

\begin{figure}
\includegraphics[width=0.5\columnwidth]{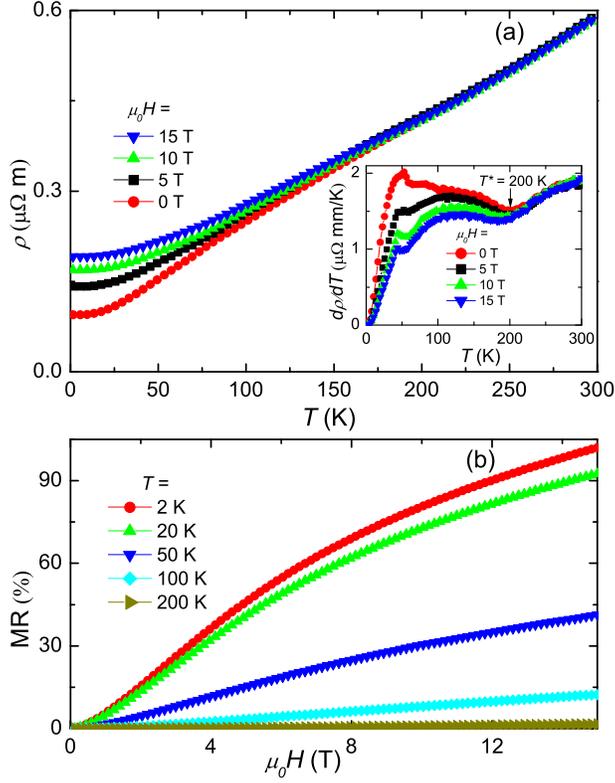}
\caption{\label{Fig. 2} \textbf{Transport properties of NbTe$_4$
at ambient pressure.} \textbf{(a)}, Plots of resistivity against
temperature under $\mu_0H$ = 0, 5, 10, and 15 T. The current is
along the $c$-axis, and the field is along the $b$-axis. Inset:
Plots of d$\rho$/d$T$ versus temperature. \textbf{(b)}, Field
dependence of MR = $[\rho(H)-\rho(0)]/\rho(0)$ under various
temperatures.}
\end{figure}

\begin{figure}
\includegraphics[width=0.5\columnwidth]{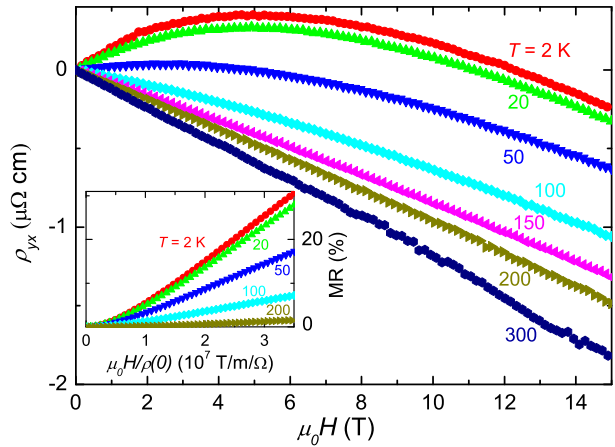}
\caption{\label{Fig. 3} \textbf{Hall resistivity and Kohler plot
for NbTe$_4$.} Field dependence of Hall resistivity ($\rho_{yx}$)
under various temperatures. The current is along the $c$-axis, and
the field is along the $b$-axis. The inset: Kohler's rule by
plotting the MR vs. $\mu_0H/\rho(0)$ from 2 K to 200 K.}
\end{figure}

\begin{figure}
\includegraphics[width=0.5\columnwidth]{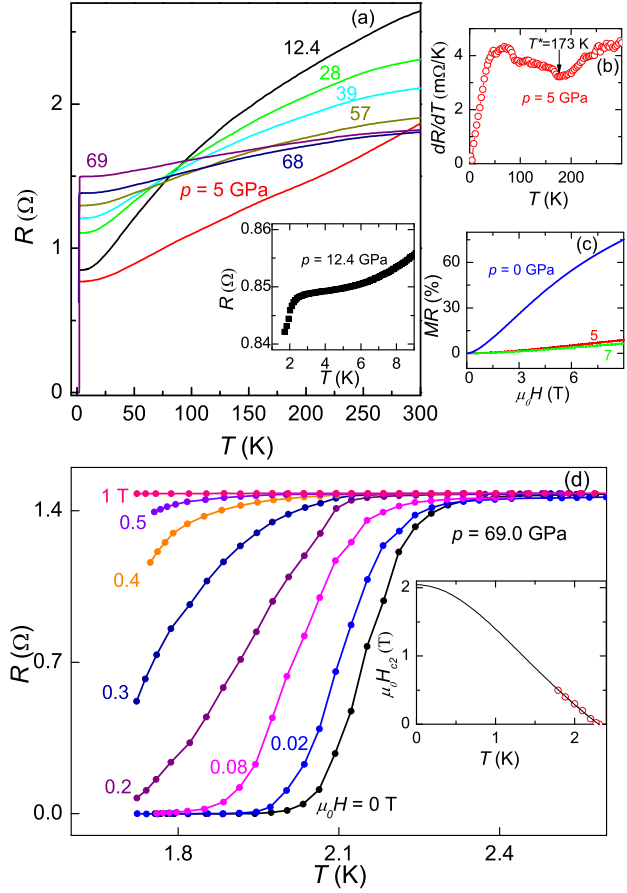}
\caption{\label{Fig. 4} \textbf{Transport properties for single
crystal NbTe$_4$ under pressure.} \textbf{(a)}, The plot of
resistance versus temperature for the pressure ranging from 5 GPa
to 69 GPa. The current is along the $c$-axis. The inset: The
enlarged plot of resistance at low temperatures under $p$ = 12.4
GPa. \textbf{(b)}, Plots of d$R$/d$T$ versus temperature under $p$
= 5 GPa. \textbf{(c)}, Field dependence of MR under $p$ = 0, 5 and
7 GPa. \textbf{(d)}, Temperature dependence of resistance for
several different magnetic fields under $p$ = 69 GPa. The inset
displays the upper critical fields as a function of temperatures.
}
\end{figure}


\begin{figure}
\includegraphics[width=0.5\columnwidth]{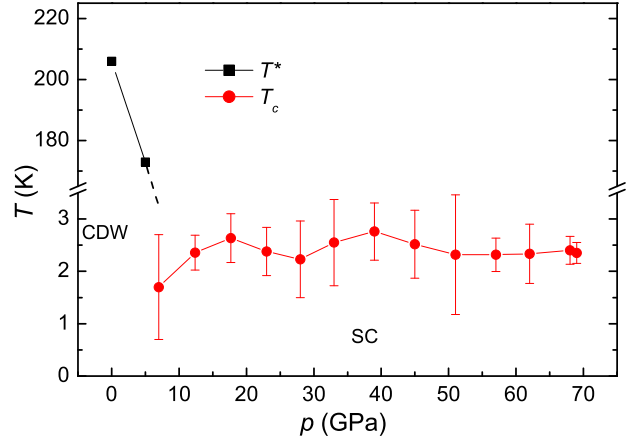}
\caption{\label{Fig. 5} \textbf{Temperature versus pressure phase diagram of NbTe$_4$.}
$T^*$ and $T_c$
denote the resistivity anomaly temperature and the superconducting transition temperature, respectively.
}
\end{figure}

\end{document}